\documentclass[runningheads]{llncs}
\usepackage{graphicx}
\usepackage{comment}
\usepackage{amsmath,amssymb} 
\usepackage{color}

\begin{document}
\pagestyle{headings}
\mainmatter

\title{Medical Image Segmentation Using a U-Net type of Architecture} 


\author{Eshal Zahra \and
Bostan Ali \and
Wajahat Siddique}
\institute{Quaid-e-Azam Medical College, Bahawalpur, Punjab 63100, Pakistan}

\maketitle
\begin{abstract}
Deep convolutional neural networks have been proven to be very effective in image related analysis and tasks, such as image segmentation, image classification, image generation, etc. Recently many sophisticated CNN based architectures have been proposed for the purpose of image segmentation. Some of these newly designed networks are used for the spcific purpose of medical image segmentation, models like V-Net, U-Net and their variants. It has been shown that U-Net produces very promising results in the domain of medical image segmentation. However, in this paper, we argue that the architecture of U-Net, when combined with a supervised training strategy at the bottleneck layer, can produce comparable results with the original U-Net architecture. More specifically, we introduce a fully supervised FC layers based pixel-wise loss at the bottleneck of the encoder branch of U-Net. The two layer based FC sub-net will train the bottleneck representation to contain more semantic information, which will be used by the decoder layers to predict the final segmentation map. The FC layer based sub-net is trained by employing the pixel-wise cross entropy loss, while the U-Net architecture is trained by using L1 loss.      
\vspace{-0.28cm}
\vspace{-0.28cm}
\end{abstract}

\section{Introduction}
From the past several years deep learning has outperformed many conventional computer vision techniques in areas such as image classification, segmentation, tracking etc.\cite{r1}, \cite{r2}, \cite{r3}, \cite{r4}, \cite{r5}, \cite{r6}. Convolutional Neural Networks (CNN) is one of the most famous deep learning architecture which is designed in 1989 \cite{r7}, but its true effectiveness came to the surface when it is trained on more powerful machines with GPUs and leveraging large amount of training data. Krizhevsky et al. \cite{r1} trained a large CNN architecture containing 8 layers and millions of parameters by using the huge ImageNet dataset with 1 million training images. From the past several years many modified and more deeper architectures of CNN has been proposed which are not only used in the medcal imaging domain but they have been widely applied to other applications as well.

Computer vision based Medical image segmentation methods can be divided into two categories, i.e, conventional medical image segmentation techniques and deep learning based methods. Some widely used conventional medical image segmentation methods include thresholding based methods \cite{r13}, \cite{r14}, \cite{r15}, region growing methods \cite{r16}, \cite{r17}, and clustering based methods \cite{r18}, \cite{r19}. Deep CNN models are mostly used for the task of image classification, however, in medical image analysis, image segmentation has its own significance, for instance image segmentation is widely used in the localization of cancerous and defected regions in MRI, CT scan and Ultrasound images. In medical image segmentation CNN models are used along with cross entropy loss as a pixel-wise measure \cite{r8}.  However, the most popular deep CNN architectures for medical image segmentation is based on an encoder-decoder architecture. The widely used models in this domain is U-Net \cite{r6} and V-Net architectures \cite{r9}. U-Net is employed for the segmentation of biological microscopy images, and since in mdeical domain the training images are not as large as in other computer vision areas, Ronneberger et al \cite{r6} trained the the U-Net model using data augmentation strategy to leverage from the available annotated images. The architecture of U-Net is consist of two main parts, i.e a contracting sub-net to encode the semantics and context information, and an expanding sub-net uses and decodes the encoded information for the generation of segmented maps. The contracting sub-net is based on down-sampling CNN blocks that extracts features
with $3 \times 3$ convolutions. The expanding sub-net is based on up-sampling CNN blocks which uses deconvolution to increase the image dimensions in spatial axis while reducing he number of channels in each image. To leverage the context information which is encoded by the intermediate layers of the contracting sub-net, the encoded feature maps are concatenated with the feature maps from the intermediate layers of deconvolutional CNN blocks of the expanding sub-net. Afterwards, $1 \times 1$ convolution is applied on the feature maps obtained from the intermediate layers of the expanding sub-net in order to produce a segmentation map in which each pixel is classified according to the corresponding semantic class of the input image. The entire U-Net architecture is trained on a dataset containing 30 transmitted
light microscopy images, and due to the efficient architectural design of this model, it won the ISBI cell tracking challenge 2015 by a significant margin. 

Similarly, V-Net \cite{r9} is another widely used image segmentation network in medical image analysis, but the main difference between this network is that it is used for 3D medical image segmentation. They proposed a loss function based on the Dice coefficient to overcome the problem of voxel imbalance in the foreground and background during the network training. V-Net is trained end-to-end on MRI voxels containing prostate information, and V-Net is trained employing the Dice coefficient to infer the segmentation for the whole volume at once. Imran et al. \cite{r10} proposed a fast segmentation method known as Progressive Dense V-net (PDV-Net) for the segmentation of pulmonary lobes from chest CT images. The PDV-net architecture contains three dense feature blocks, which processes the entire CT volume in order to generate the segmentation information in an automatic manner. As opposed to existing medical image segmentation methods which requires prior information, PDV-Net eliminates the need for any user interaction in the form of providing prior information. Similarly, \cite{r11} implements a 3D-CNN encoder for lesion segmentation which combines the advantages of U-Net and CEN \cite{r12}. The 3D-CNN network is consist of two branches, a conventional convolutional branch and a deconvolutional branch. The convolutional branch is based on convolutional and pooling layers and the deconvolutional branch contains deconvolutional and unpooling layers.

In this paper, we present an architecture, which is quite similar to the aforementioned networks, but the main difference between our proposed method and the existing medical image segmentation techniques discussed in the previous paragraph is that we combine the advantages of supervised learning with the self-supervised training strategy of a typical U-Net architecture. We argue that, by explicitly, providing the supervisory signal at the bottleneck layer of the encoder part of U-Net, the encoder or the contracting branch can encode more effective features as compared to using self-supervised training approach.
 
 \begin{figure}[t]
    \centering
    \includegraphics[width=10.5cm,, height=5cm]{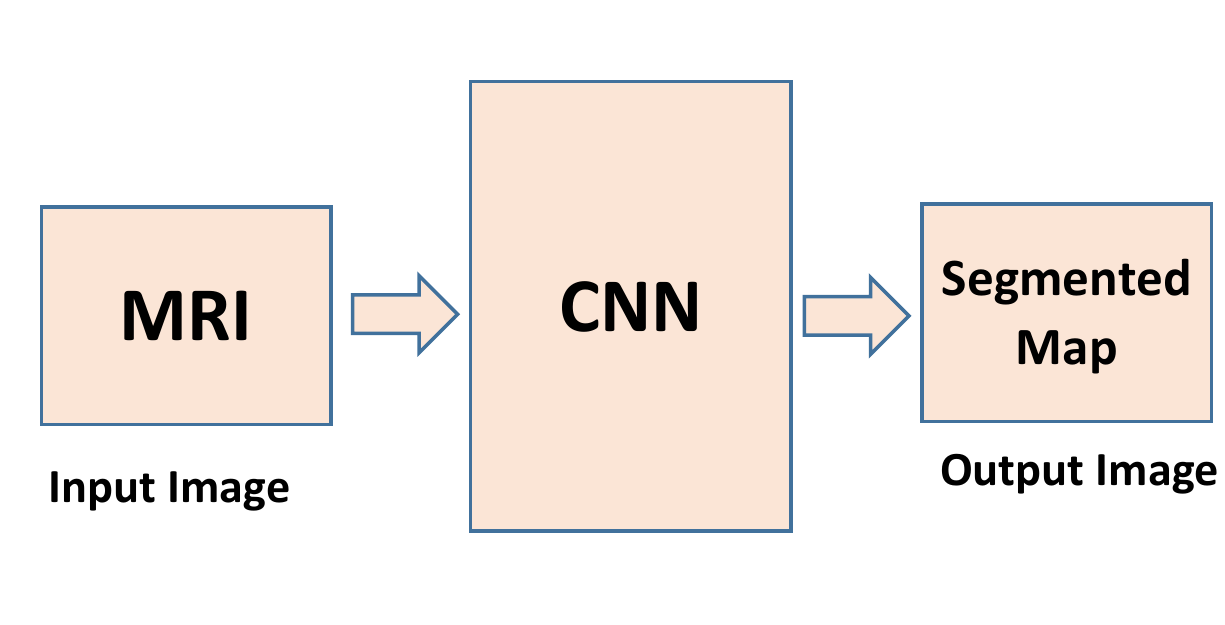}
    \vspace{-0.58cm}
    \caption{The main idea: Our CNN network takes an input medical image and passes it through its intermediate layers and produces a segmentation map using its decoder part}
    \label{fig:1}
    \vspace{-0.5cm}
\end{figure}

\section{Network Architecture}
\begin{figure*}[t]
\centering
\includegraphics[width=12cm,, height=9cm]{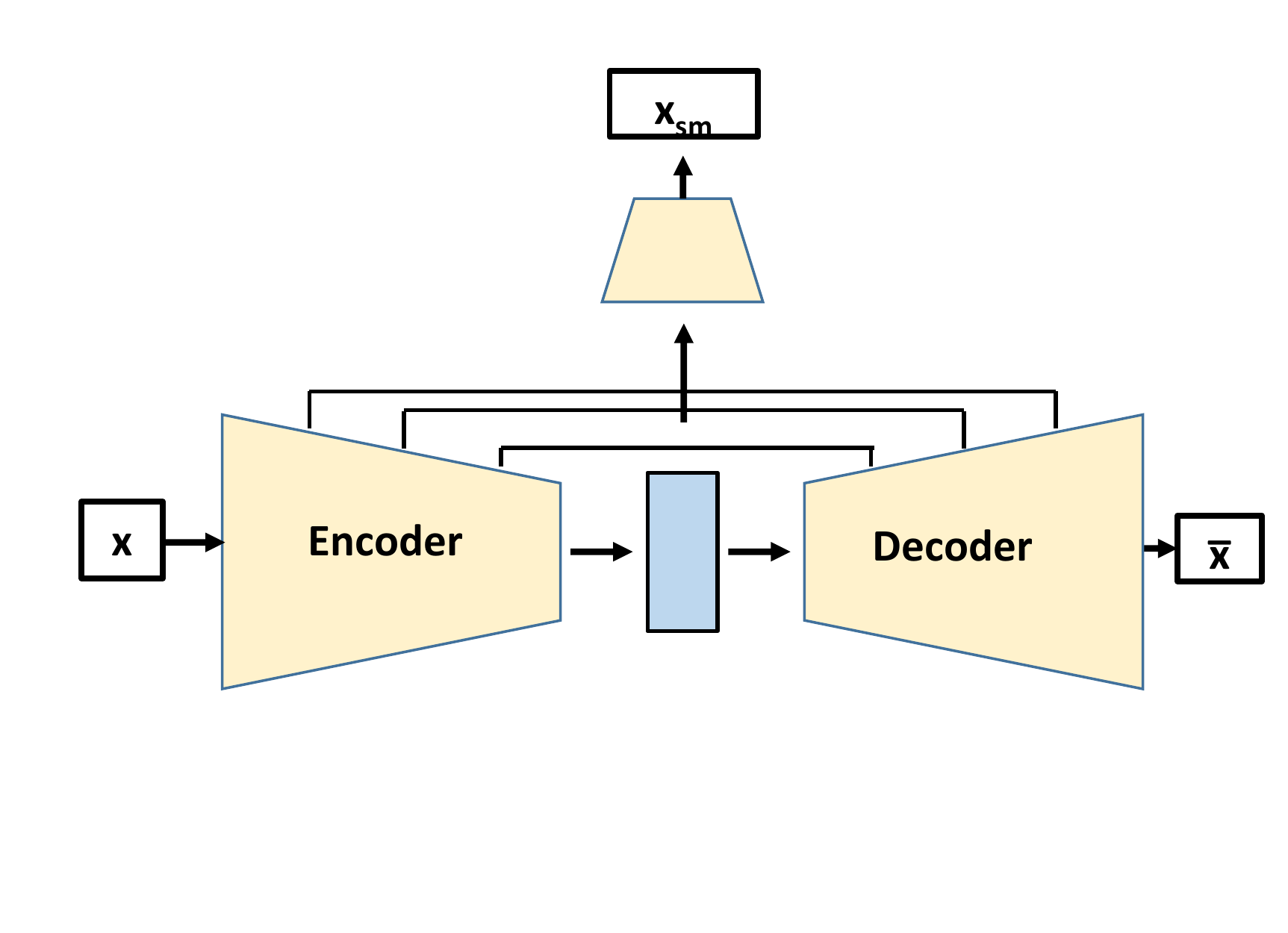}
\vspace{-0.25cm}
\caption{The over-all architecture of our proposed network: An input image, such as an MRI or CT-scan image is fed to the CNN based network which extracts context information in its intermediate layers and this encoding of the context information is enhanced by using a bottleneck training layer. The decoder part of our network then uses the encoded information to generate the segmented map using skip connections from encoder layers to intermediate layers of decoder }
\vspace{-0.35cm}
\label{fig:2}
\vspace{-0.25cm}
\end{figure*}

The overall framework of our proposed technique is shown in Figure \ref{fig:2}. The network is consist of three parts, i.e 1) an encoder part, 2) bottleneck training part, and 3) the decoder part. The encoder part is based on
the typical design of a convolutional neural network which contains convolutional blocks with $3 \times 3$ filters. Each convolutional block is followed by a a rectified linear unit (ReLU) and a down-sampling layer having 2x2 max pooling operation with stride 2. The down-sampling layer reduces the size of the input image spatially, while it increases the number of channels of the feature maps to encode more useful information. The bottleneck training part is consist of two fully connected layers to predict the ground-truth segmentation map by using linear transformation as the input image and the predicted segmentation maps are registered. The decoder part of the network is designed based on the up-sampling deconvolutional blocks. We use $2 \times 2$ up-convolutions to increase the size of the feature maps in the intermediate de-convolutional layers. Following the skip-connection architecture of U-net we concatenate feature maps from the encoder layers to the corresponding layer in the decoder network. We then use $3 \times 3$ convolutional filters followed by a ReLU to incorporate non-linearity in this branch of the network.   
\section{Results and Discussion}
The proposed CNN based method is evaluated using the criteria of sensitivity and specificity, defined by the following formulas:

\begin{align}
Sensitivity = {}&N_{tp}/N_p\\
Specificity = {}&N_{tn}/N_n\\
Accuracy = {}&N_{tp}+N_{tn}/N_{tp}+N_{tn}+N_{fn}+N_{fp}
\label{eq:2}
\end{align}
Table \ref{table:1} shows the specificity, sensitivity and accuracy obtained by training and validating our proposed model on MRI and CT-scan images.  

\begin{table}[t!]
\begin{center}
\begin{tabular}{|l| c| c|c|} 
 \hline
 Data&Specificity&Sensitivity&Accuracy\\
 \hline
 \hline
 MRI& 0.926  & 0.939 & 0.913 \\
  \hline
 CT Scan& 0.961  & 0.972& 0.976\\
 \hline
\end{tabular}
\end{center}
\vspace{-0.3cm}
\caption{Oulu-CASIA: Accuracy for six expressions classification.}
\vspace{-0.1cm}
\label{table:1}
\vspace{-0.55 cm}
\end{table}

\section{Conclusion}
In this paper we have presented a U-Net type of architecture, which is based on convolutional neural networks for medical image segmentation. Our proposed network has three parts, i.e, 1) an encoder part, 2) a bottleneck learning layer and a 3) decoder part of the network. The encoder part encodes the context information from the input image in the intermediate layers using CNN filters followed by non-linearity of RELU. The bottleneck layer is used to enhance the feature extraction capability of the encoder part by using a fully supervised linear transformation based on fully connected layers. The FC layers in the bottleneck part of the network is used to predict the ground truth segmentation map using a linear transformation. The decoder part of our network is based on deconvolutional blocks which increases the spatial dimensions of feature maps and reduces the channels of the feature maps in its intermediate layers. To take full advantage of the encoded information in the intermediate layers of the encoder and to prevent the loss of information, we add skip connections, connecting the intermediate layers of the encoder with the intermediate layers of the decoder. Experimental results show that the proposed technique produces promising results on MRI and CT scan images.

%
%
\bibliographystyle{splncs04}
\bibliography{egbib}
\end{document}